# A Balanced Energy Consumption Clustering Algorithm for Heterogeneous Energy Wireless Sensor Networks

Xiaofu Ma, Yu Fang, Xingzhen Bai

Dept. of Computer Science & Engineering
Tongji University
Shanghai, P.R.China
maxiaofu2008@163.com

*Abstract*—**In this paper, a balanced energy consumption clustering algorithm (BECC) is proposed. This new scheme is a cluster-based algorithm designed for heterogeneous energy wireless sensor networks. A polarized energy factor is introduced to adjust the probability with which each node may become a cluster head in the election of the new clustering scheme. Under the condition that the expected number of cluster heads in the network preserves the theoretical optimal number, BECC makes sure that nodes with higher residual energy will become cluster heads with higher probabilities while nodes with lower residual energy will not become cluster heads. Simulation results show that this new scheme provides longer lifetime than the classical clustering algorithms including LEACH and other improved algorithms in heterogeneous networks, and BECC also reaches larger amount of messages received at the sink.**

## I. INTRODUCTION

Wireless sensor networks are networks of numerous tiny battery powered sensor nodes with low cost and limited energy. Through the cooperation with each other, sensor nodes sense the environment, collect sensed data and send their reports towards a sink. The sink is a processing center, and typically serves as a gateway to some other networks [1]. For the very limited power of sensor nodes and the difficulty of the battery renewal, it is necessary to design energy-efficient protocols to maximize the network lifetime.

In order to reduce the energy consumption of wireless sensor networks, many energy-efficient routing protocols apply hierarchical or cluster-based architecture which is an excellent technique with special advantages related to scalability and efficient communication [2-4]. Most of these schemes assume that all the nodes in the network are homogenous, which means nodes are equipped with the same amount of energy. However, it is of great significance to design algorithms for heterogonous environment since the diversity of the initial energy among all the nodes and the unbalanced consumption caused by various kinds of random events during the network operation time may lead to heterogeneous cases.

In this paper, we propose a balanced energy consumption clustering algorithm (BECC). BECC is based on LEACH （Low-Energy Adaptive Clustering Hierarchy） protocol [2], and introduces a polarized energy factor as a reference to adjust the probability with which each node will become a cluster head. It makes sure that nodes with higher residual energy will become cluster heads with higher probabilities while the low will not become cluster heads so as to balance the energy consumption. BECC also guarantees that the expected number of cluster heads, during every period of the network operation time, preserves the theoretical optimal number. We show by simulation that BECC significantly outperforms classical clustering algorithms including LEACH [2], LEACH-E [5] and SEP [6] in terms of the network lifetime and the amount of messages received at the sink in heterogeneous energy wireless sensor networks.

The remainder of this paper is organized as follows. Section 2 reviews related works. Section 3 describes the network model, assumptions and wireless radio model. Section 4 presents the detailed design of BECC. Section 5 reports the result of BECC effectiveness and performance via simulations and compares it with classical clustering algorithms. Section 6 concludes the paper.

## II. RELATED WORK

Low-energy adaptive clustering hierarchy (LEACH) [2] is one of the most popular distributed cluster-based routing protocols in homogeneous wireless sensor networks. By rotating the cluster head role uniformly and periodically among the nodes, each node tends to expend the same energy over time. LEACH divides the operation time into slots called "round". Each round is generally separated into two phases, the set-up phase and the steady-state phase. In the set-up phase, $p_{opt}N$ cluster heads will be elected approximately, where $p_{opt}$ is the predetermined optimal percentage of cluster heads, and $N$ is the total number of nodes in the network. Each node $s_i$ decides whether or not to become a cluster head for the current round according to the following threshold

$$T(s_i) = \begin{cases} \dfrac{p_{opt}}{1 - p_{opt}(r \bmod \frac{1}{p_{opt}})}, & if \quad s_i \in G \\ 0, & otherwise \end{cases} \qquad (1)$$

where $r$ is the current round, and $G$ is the set of nodes that have not been cluster heads in the last $(r \bmod (1/p_{opt}))$ rounds.

In LEACH, the optimal number of cluster heads is estimated to be about 5% of the total number of nodes. Each node which has elected itself as a cluster head for the current round broadcasts an advertisement message to the rest of the nodes in the network. All the non-cluster head nodes, after receiving the advertisement messages, decide on the cluster to which they will belong for this round. This decision is based on the received signal strength of the advertisement messages. And then each non-cluster head node transmits a join-request message (Join-REQ) to its chosen cluster head. After receiving all the messages from the nodes that would like to be included in the cluster, the cluster head creates a TDMA (Time Division Multiple Address) schedule and assigns each node a time slot when it can transmit.

During the steady-state phase, non-cluster head nodes transmit the sensed date to cluster heads according to their own allocated transmission time. The cluster head, after receiving all the data, aggregate it before sending it to the sink.

LEACH is typically designed for homogeneous wireless sensor networks, and the authors of [5] proposed a solution to meet the needs of energy-efficient performance in heterogeneous networks. This scheme improves LEACH and is called LEACH-E in this paper. The residual energy of each node and the total energy of the network are taken into consideration when each node decides whether or not to become a cluster head. But since it is very hard to acquire the global energy information in wireless sensor networks, the scalability of this protocol will be influenced.

SEP is proposed by [6], which is feasible when the network is two-level heterogeneous. Only two types of initial energy among the nodes exist in a two-level heterogeneous network. In multi-level heterogeneous networks where the initial energy of nodes is randomly distributed in a certain range, SEP is not suitable.

In [7], the authors extend SEP algorithm for multi-level heterogeneous networks, which is called SEP-M. And the authors also introduced a new cluster-based scheme DEEC for both multi-level and two-level heterogeneous networks with better performance.

CODA is proposed by [8] in order to relieve the unbalance of energy depletion caused by different distance from the sink. However, the work of CODA relies on global information of node positions, and thus it is not scalable.

HEED [9] is an energy-efficient cluster-based algorithm which periodically selects cluster heads according to a hybrid of the node's residual energy and a secondary parameter such as node degree.

In [10], a new algorithm is proposed to exploit the redundancy properties of the wireless sensor networks. And it also changes the inter cluster communication pattern depending on the energy condition of the high energy nodes during the lifetime of the heterogeneous networks.

In [11], the authors introduce an energy-efficient heterogeneous clustered scheme EEHC based on weighted election probabilities of each node to become a cluster head according to the node's initial energy relative to that of other nodes in the network.

## III. MODEL DESCRIPTION

### *A. Network Model and Assumption*

The network model and assumptions in this paper are the same as [2]. Assume $N$ sensor nodes are distributed over a square area whose side length is $M$ meters. The sensor nodes monitor the environment and transmit the sensed data to the sink periodically. All the nodes are static or slightly move and are location-unaware. Nodes' radio transceivers are capable of changing the transmission power continuously to achieve different transmission ranges.

### *B. Wireless Radio Model*

In this paper, we use similar radio dissipation model as proposed in [5]. To transmit a $k$ bit message a distance $d$, the radio expends

$$E_S(k,d)=\begin{cases} kE_{elec}+k\varepsilon_{fs}d^2 & d<d_0 \\ kE_{elec}+k\varepsilon_{mp}d^4 & d\ge d_0 \end{cases} \tag{2}$$

where $E_{elec}$ is the energy dissipated per bit to run the transmitter or the receiver circuit, $\varepsilon_{fs}$ and $\varepsilon_{mp}$ depend on the transmitter amplifier model we use, $d_0=\sqrt{\varepsilon_{fs}/\varepsilon_{mp}}$ is a constant.

And to receive this message, the radio expends

$$E_r(k)=kE_{elec} \tag{3}$$

To aggregate $m$ packets with the length of $k$, the energy consumption is

$$E_f(k)=mkE_{DA} \tag{4}$$

where $E_{DA}$ is the energy consumption of aggregating data with the length of 1 bit.

## IV. BECC ALGORITHM

In this section, we describe our clustering algorithm BECC. It improves the election pattern of LEACH and is designed for heterogeneous energy wireless sensor networks.

Similarly to LEACH, BECC divides the network operation time into several rounds. Each round is also separated into the set-up phase and the steady-state phase. In contrast to LEACH, each node, during the current round, acquires its polarized energy factor in a distributed way. The polarized energy factor is used to determine the node's threshold of being elected as a cluster head in the coming round.

In the set-up phase of BECC, when a non-cluster head node transmits a Join-REQ to its chosen cluster head as described in LEACH protocol, the node need to piggyback its residual energy information together with the Join-REQ to its chosen cluster head.

Cluster heads, after receiving these messages, have enough information to calculate the polarized energy factor for each node in their corresponding clusters. The definition and calculation of the relative energy factor, the polarized energy factor and other parameters are as follows.

If the node number of the cluster to which node $s_i$ belongs is $n$, some parameters of $s_i$ are defined as

$$q_{rel}(s_i) = \frac{E_i(r)}{E_{total}(r)} n \tag{5}$$

$$f_{gt1}(s_i) = \begin{cases} 1, & if \quad q_{rel}(s_i) \geq 1 \\ 0, & otherwise \end{cases} \tag{6}$$

$$f_{lt1}(s_i) = \begin{cases} 1, & if \quad q_{rel}(s_i) < 1 \\ 0, & otherwise \end{cases} \tag{7}$$

where $E_i(r)$ is the residual energy of $s_i$ in $r$ round, and $E_{total}(r) = \sum_{i=1}^{n} E_i(r)$ is the total energy of the cluster to which $s_i$ belongs.

$q_{rel}(s_i)$ is a relative residual energy metrics of $s_i$, $f_{gt1}(s_i)$ identifies whether the residual energy of $s_i$ is greater than or equal to the average energy, and $f_{lt1}(s_i)$ identifies whether the residual energy of $s_i$ is less than the average energy.

Recall that $p_{opt}$ is optimal percentage of cluster heads and then the threshold of being a cluster head for $s_i$ in the coming round is set to

$$T(s_i) = p_{opt} q_{pol}(s_i) \tag{8}$$

where $q_{pol}(s_i)$ defined as follows is the polarized energy factor of $s_i$

$$q_{pol}(s_i) = f_{gt1}(s_i) \left( q_{rel}(s_i) + \frac{q_{rel}(s_i) \sum_{i=1}^{n} q_{rel}(s_i) f_{lt1}(s_i)}{\sum_{i=1}^{n} q_{rel}(s_i) f_{gt1}(s_i)} \right) \tag{9}$$

During the same set-up phase, each cluster head piggybacks the polarized energy factor $q_{pol}(s_i)$ together with the TDMA schedule to the nodes in its cluster. Since the bit lengths of the residual energy and the polarized energy factor are very short, this "piggyback" process may affect the energy consumption as well as the network lifetime very slightly.

The steady-state phase in BECC is the same as that in LEACH when the sensed data transmission starts.

BECC is designed to achieve energy-efficient performance in heterogeneous wireless sensor networks. It is easy to learn from the expressions of $q_{pol}(s_i)$ and $T(s_i)$ that the value of $q_{pol}(s_i)$ is equal to zero so that $s_i$ will not be elected as a cluster head when the residual energy of $s_i$ is less than the average energy. On the other hand, the expressions show that the greater the value of $q_{rel}(s_i)$, the greater the value of $q_{pol}(s_i)$ when the energy of $s_i$ is above the average energy, thus it guarantees that a node with higher residual energy will become a cluster head with higher probability.

The expected number of cluster heads in the coming round elected from the cluster to which $s_i$ belongs in the current round is

$$\begin{aligned} E(\#CH) &= \sum_{i=1}^{n} p_{opt} q_{pol}(s_i) \\ &= p_{opt} \sum_{i=1}^{n} \left( \frac{f_{gt1}(s_i) q_{rel}(s_i) \left( \sum_{i=1}^{n} q_{rel}(s_i) f_{gt1}(s_i) + \sum_{i=1}^{n} q_{rel}(s_i) f_{lt1}(s_i) \right)}{\sum_{i=1}^{n} q_{rel}(s_i) f_{gt1}(s_i)} \right) \\ &= p_{opt} \sum_{i=1}^{N} \left( \frac{f_{gt1}(s_i) q_{rel}(s_i) \sum_{i=1}^{N} \left( \frac{E_i(r)}{E_{total}(r)} N \right)}{\sum_{i=1}^{N} q_{rel}(s_i) f_{gt1}(s_i)} \right) \\ &= p_{opt} \sum_{i=1}^{N} \left( \frac{f_{gt1}(s_i) q_{rel}(s_i) \frac{\sum_{i=1}^{N} E_i(r)}{E_{total}(r)} N}{\sum_{i=1}^{N} q_{rel}(s_i) f_{gt1}(s_i)} \right) \\ &= p_{opt} n \end{aligned}$$

Let $N$ be the total number of nodes in the network, $m$ be the cluster number in the current round, and $n_k$ be the number of nodes in each cluster where $k = 1, 2, ..., m$, then the expected number of cluster heads in the network is

$$E(\#CH) = \sum_{k=1}^{m} p_{opt} n_k = p_{opt} \sum_{k=1}^{m} n_k = p_{opt} N$$

Since $p_{opt} N$ is the theoretical optimal number of cluster heads for the network, the total energy consumption of the network is minimized.

## V. SIMULATION RESULTS

To evaluate the performance of our algorithm, we do the simulation using MATLAB. A heterogeneous wireless sensor network with 200 nodes randomly distributed in a field with dimensions $500m \times 500m$ is studied. For simplicity, we assume the sink is located in the center of the network. The impact caused by random factors such as signal collision and wireless channel interference is ignored in the process of simulation. Both $q_{rel}(s_i)$ and $q_{pol}(s_i)$ are set to one for each node before the first round. The parameters used in the simulations are summarized in Table I.

TABLE I. SIMULATION PARAMETER VALUE

| Parameter | Value |
|---|---|
| *Network grid* | (0,0)-(500,500) |
| *Node number* | 200 |
| $E_{elec}$ $nJ/b$ | 50 |
| $E_{DA}$ $nJ/bit/signal$ | 5 |
| $\varepsilon_{fs}$ $pJ/bit/m^2$ | 10 |
| $\varepsilon_{mp}$ $pJ/bit/m^4$ | 0.0013 |
| *Message size* (*bit*) | 4000 |

### *A. Analyses of Two-level Heterogeneous Networks*

In a two-level heterogeneous energy wireless sensor network, let $\lambda$ be the fraction of the number of the nodes which are equipped with $\alpha$ times more energy than the others. These powerful nodes are called advanced nodes, and the others are normal nodes whose initial energy is $E_0$. In addition, the stability period which reflects the network lifetime is defined as a time period from the beginning of the network operation time until the death of the first node.

To validate the performance of BECC in two-level heterogeneous networks, we study, by varying $\lambda$ and $\alpha$, the stability period using LEACH, LEACH-E, SEP and BECC.

Figure 1 shows the stability period versus $\lambda$ which varies from 0.1 to 0.9, and figure 2 shows the stability period versus $\alpha$ which varies from 0.5 to 4.5 in two-level heterogeneous networks.

We observe that, as expected, the stability period using LEACH fluctuates slightly as $\alpha$ or $\lambda$ increases. It shows that LEACH cannot make full use of the increased network energy, and thus is not suitable in heterogeneous networks. This is because each node in LEACH is given equal opportunity to become a cluster head, and nodes with low initial energy may die very quickly while the residual energy of nodes with high initial energy still remains very high. Since the death of low energy nodes determines the stability period of the network, the stability period using LEACH in heterogeneous environment is very short.

By contrast, the curves of LEACH-E, SEP and BECC tend to go upward obviously. This is because the heterogeneity of nodes is taken into consideration in these three algorithms. And thus the stability periods using these three algorithms increase along with the incremental "extra" energy caused by the variation of $\alpha$ and $\lambda$. Compared with that using LEACH, LEACH-E and SEP, the stability period using BECC increases by 98%, 22% and 58% respectively when $\lambda$ varies. And it increases by 127%, 57% and 31% respectively when $\alpha$ varies.

### B. Analyses of Multi-level Heterogeneous Networks

In a multi-level heterogeneous network, the initial energy of nodes is randomly distributed in a closed interval $[E_{min}, E_{max}]$, where $E_{min}$ is the lower bound of the energy and $E_{max}$ is the upper bound of the energy. Let the total energy of each multi-level heterogeneous network be the same to avoid the impact of different total energy caused by randomness.

To validate the performance of BECC in multi-level heterogeneous networks, we consider the case when the initial energy of nodes in the network is randomly distributed in a closed interval $[1J, 5J]$. Figure 3, 4 and 5 show how the standard deviation of residual energy of nodes alive in the network, the number of nodes alive in the network and the number of messages received at the sink change over time using LEACH, LEACH-E, SEP-M and BECC.

We observe from figure 3 that the standard deviation using LEACH tends to increase at first, which means LEACH has no effect on balancing energy consumption. The downward tendency of the LEACH curve is due to the decreasing number of nodes alive. The standard deviation curves of LEACH-E, SEP and BECC monotonically decrease over time shows that the effect of balancing energy consumption can be achieved by these three algorithms. It is clearly that BECC outperforms all the other algorithms since the curve of BECC goes down fastest.

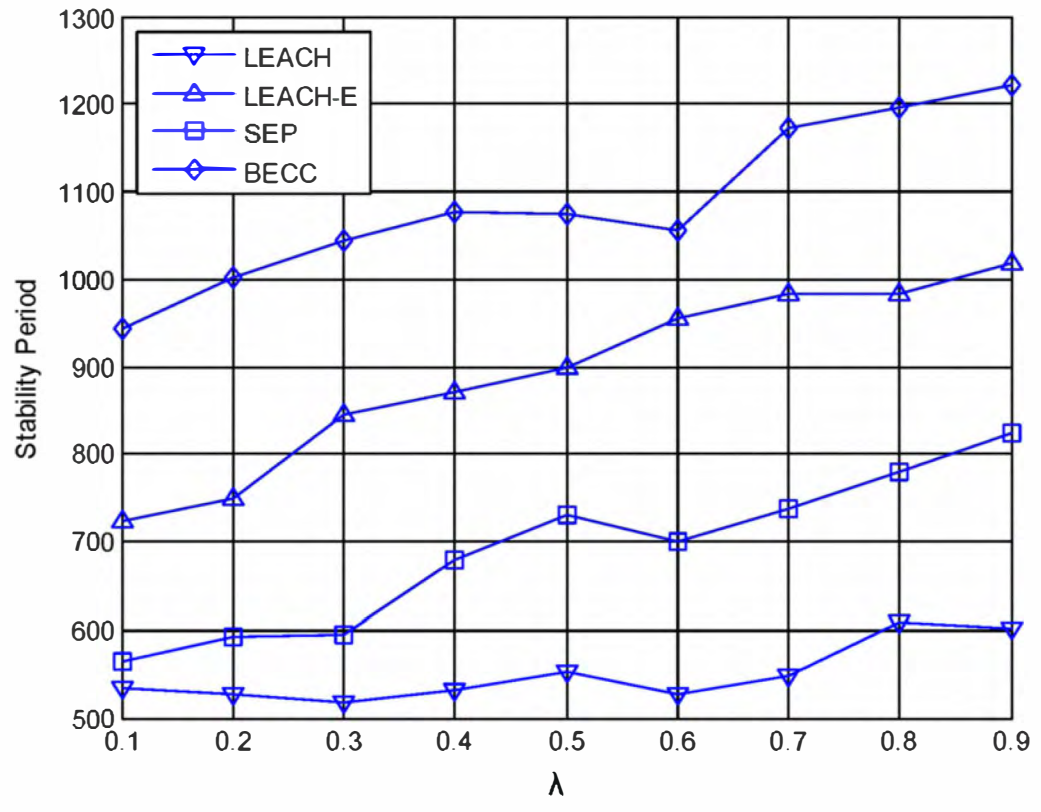

Figure 1. Stability period when $\lambda$ is varying

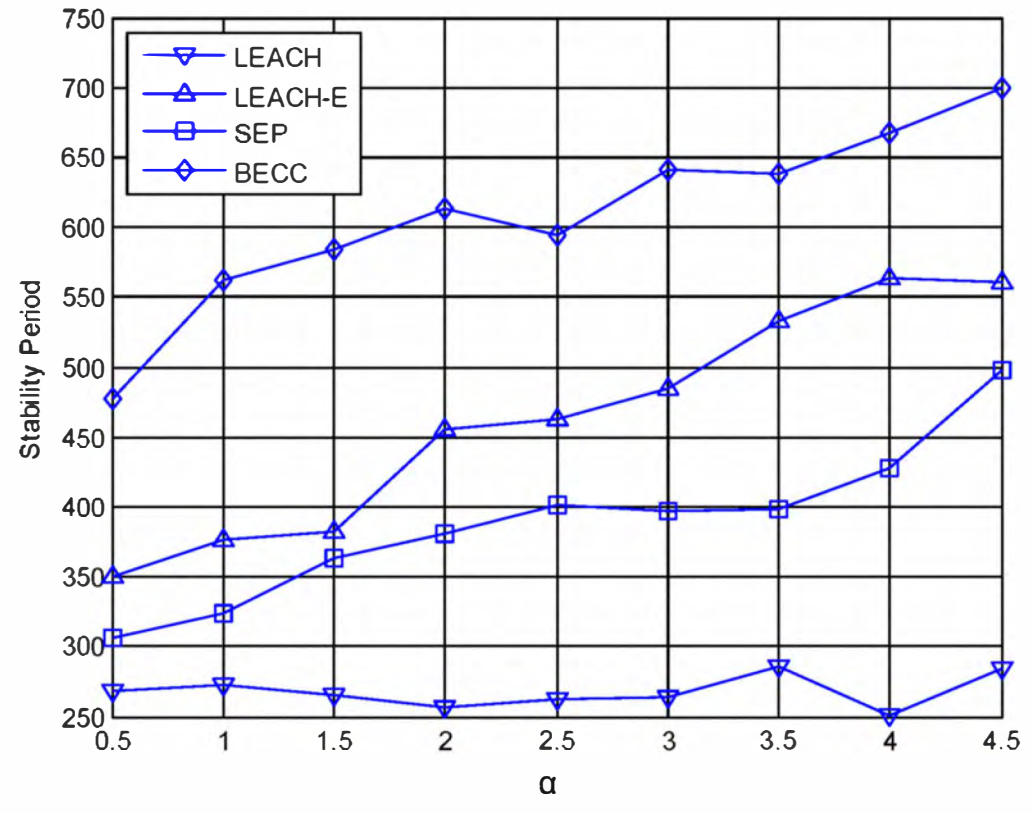

Figure 2. Stability period when $\alpha$ is varying

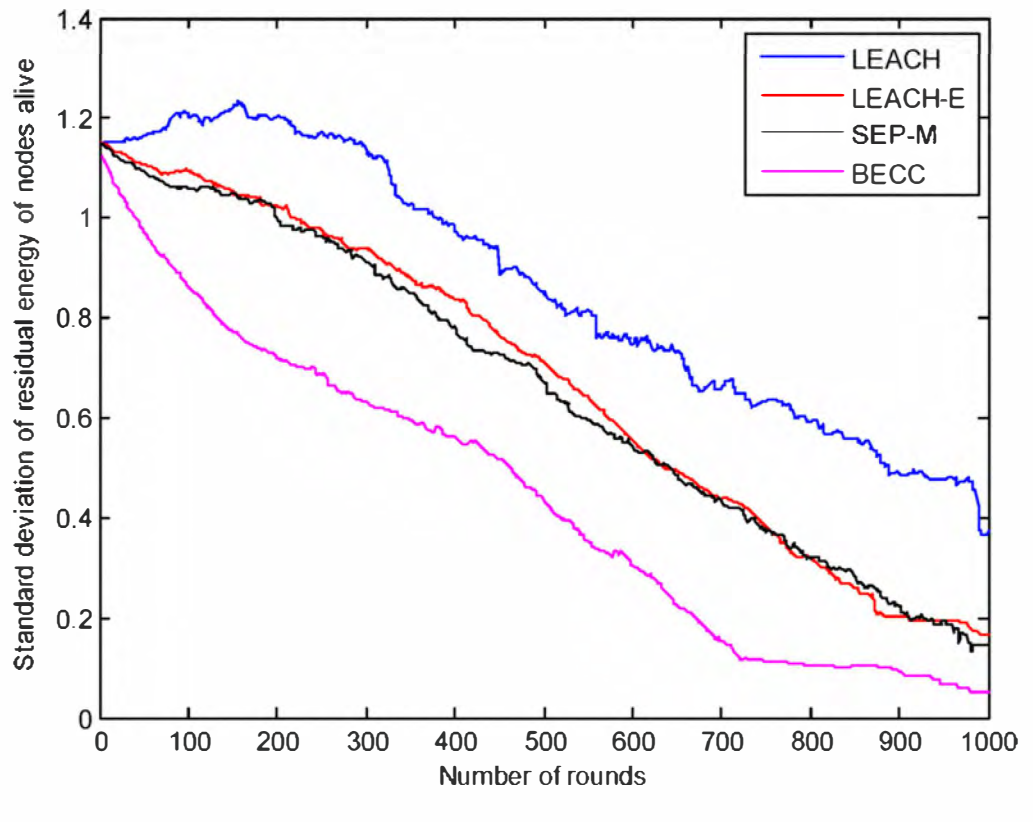

Figure 3. Standard deviation of residual energy of nodes alive over time

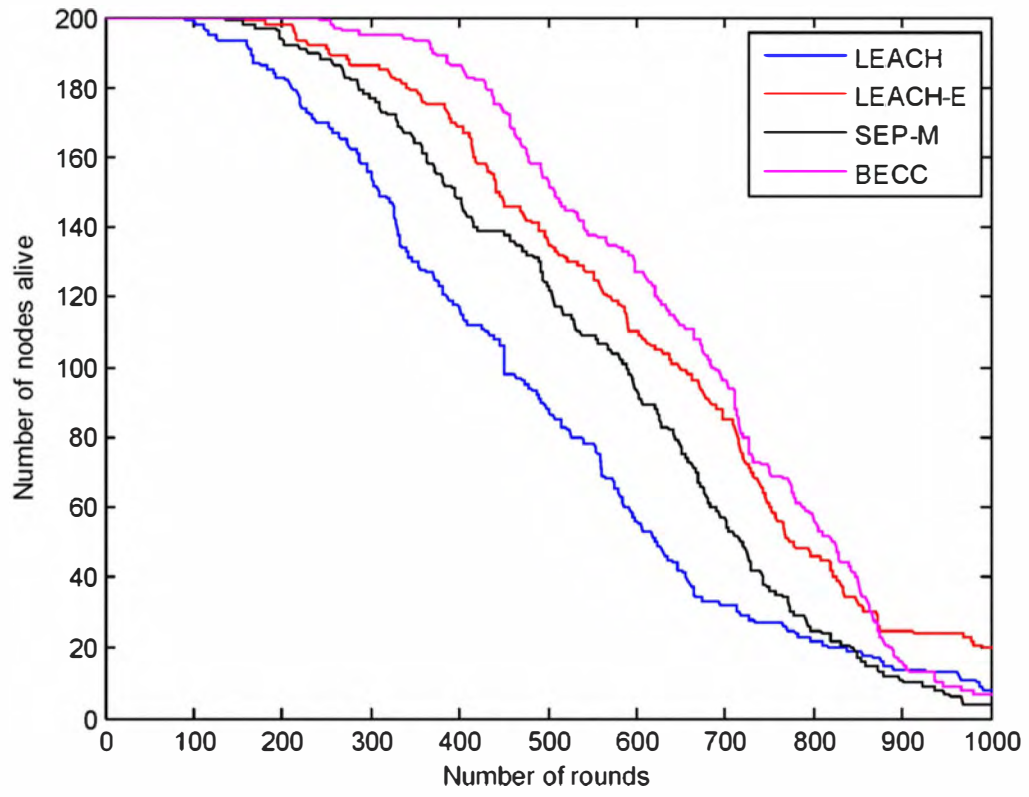

Figure 4. Number of nodes alive over time

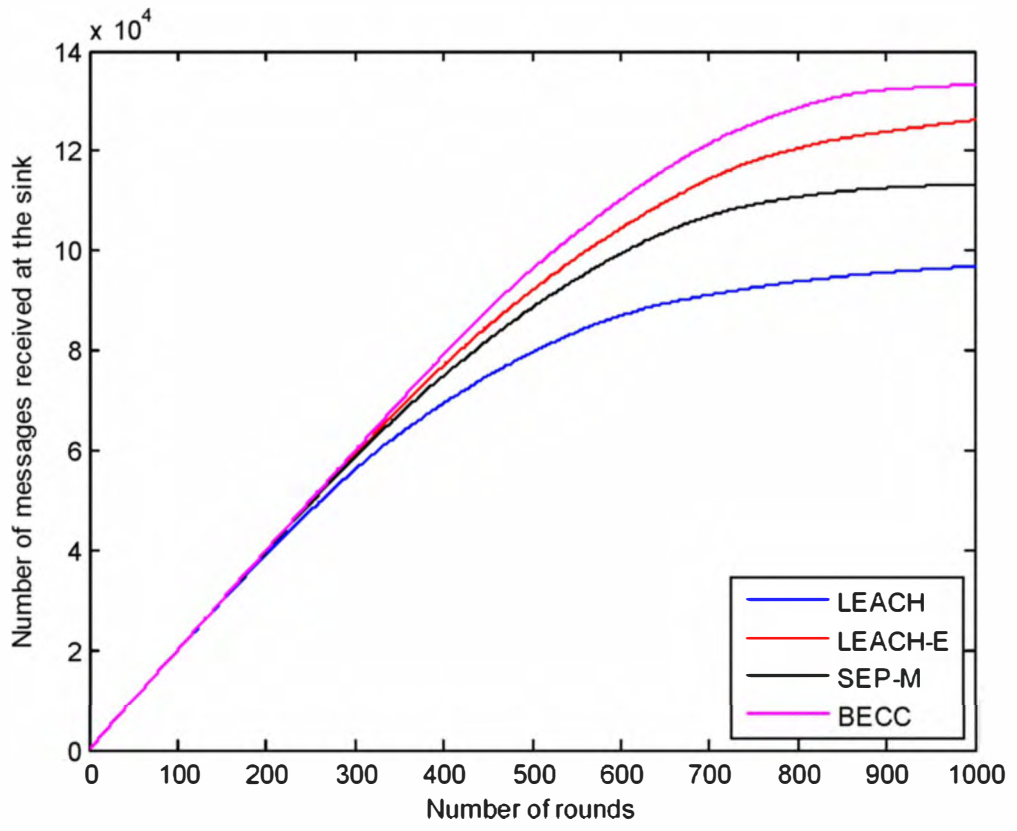

Figure 5. Number of messages received at the sink over time

Figure 4 shows that the stability periods using LEACH, LEACH-E, SEP-M and BECC are 91, 152, 137 and 244, respectively, in the multi-level heterogeneous network. Meanwhile, we observe that the number of nodes alive using BECC is greater than that using LEACH, LEACH-E and SEP-M in each round from the beginning of the network operation until more than 80% of the nodes dies. It is thus clear that BECC prolong the lifetime effectively in multi-level heterogeneous networks.

When it comes to the number of messages received at the sink, figure 5 shows, using BECC, the measured value which reflects both of the network throughput and the amount of effective data is highest as expected due to its extended stability and lifetime.

## VI. CONCLUSIONS AND FUTURE WORK

In this paper, a balanced energy consumption clustering algorithm (BECC) is proposed for heterogeneous wireless sensor networks. It inherits the advantages of LEACH, since BECC does not require global knowledge of energy in each election round, the data processing and calculation can be accomplished in a distributed way and it does not require the exact position of each node in the field during the whole operation time. BECC use a polarized energy factor to adjust the threshold of each node being a cluster head, in order to balance the energy consumption. Simulation results demonstrate that BECC significantly outperforms classical clustering algorithms in terms of balancing energy consumption, prolonging network lifetime and providing larger amount of messages received at the sink.

We are currently extending BECC to design a multi-hop clustering and routing protocol to achieve better performance in heterogeneous energy wireless sensor networks.

## ACKNOWLEDGMENT

This research is supported in part by National Defense Basic Research Program of China's Ministry of Education (Grant No.xxxxxxxxxxx).